# An Analysis of Introductory Programming Courses at UK Universities

Ellen Murphy[a], Tom Crick[b], and James H. Davenport[c]

a   Institute for Mathematical Innovation, University of Bath, UK
b   Department of Computing & Information Systems, Cardiff Metropolitan University, UK
c   Department of Computer Science, University of Bath, UK

**Abstract**   *Context:* In the context of exploring the art, science and engineering of programming, the question of which programming languages should be taught first has been fiercely debated since computer science teaching started in universities. Failure to grasp programming readily almost certainly implies failure to progress in computer science.
*Inquiry:* What first programming languages *are* being taught? There have been regular national-scale surveys in Australia and New Zealand, with the only US survey reporting on a small subset of universities. This the first such national survey of universities in the UK.
*Approach:* We report the results of the first survey of introductory programming courses (N = 80) taught at UK universities as part of their first year computer science (or related) degree programmes, conducted in the first half of 2016. We report on student numbers, programming paradigm, programming languages and environment/tools used, as well as the underpinning rationale for these choices.
*Knowledge:* The results in this first UK survey indicate a dominance of Java at a time when universities are still generally teaching students who are new to programming (and computer science), despite the fact that Python is perceived, by the same respondents, to be both easier to teach as well as to learn.
*Grounding:* We compare the results of this survey with a related survey conducted since 2010 (as well as earlier surveys from 2001 and 2003) in Australia and New Zealand.
*Importance:* This survey provides a starting point for valuable pedagogic baseline data for the analysis of the art, science and engineering of programming, in the context of substantial computer science curriculum reform in UK schools, as well as increasing scrutiny of teaching excellence and graduate employability for UK universities.

# The Art, Science, and Engineering of Programming

An Analysis of Introductory Programming Courses at UK Universities

# 1 Introduction

For many years – and increasingly at all levels of compulsory and post-compulsory education – the choice of programming language to introduce the "art" [22], "science" [17] and "discipline" [13] of computer programming via key programming principles, constructs, syntax and semantics has been regularly revisited. Even in the context of what are perceived to be the most challenging introductory topics in computer science degrees, numerous key themes across programming frequently appear [10].

So what makes a good first programming language? Perhaps more importantly, how are we defining "good"? The issues surrounding choosing a first language [19, 21] – and a search of the ACM Digital Library identified a number of papers of the form "*X as a first programming language*", going as far back as the 1970s [16] – appear to be legion, especially with wider discussions of precisely what we aim to achieve from teaching programming [15, 35, 41, 1, 4]; from the potential impact on students' grades and attainment [37, 3, 31, 25, 20], addressing gender and diversity issues [2, 38], to a renewed focus on developing transferable computational thinking and problem solving skills [29, 42, 39]. There is a belief that programming – as opposed to, say analysis of algorithms, a closely related theoretical skill – is fundamentally a craft that needs immersion and practice [15, 28]. It appears that decades of research on the teaching of introductory programming has had limited effect on classroom practice [30]; although relevant research exists across several disciplines including education and cognitive science, disciplinary differences have often made this material inaccessible to many computing educators. Furthermore, computer science instructors have not had access to comprehensive surveys of research in this area [26, 30].

However, in Australia and New Zealand there have been longitudinal data collections [33, 24, 23] surveying the teaching of introductory programming courses in universities. Surprisingly, such surveys have not been conducted elsewhere on this scale (in the USA, [18] only covers the "top 39" universities), and this paper reports the findings from running the first such similar survey in the UK.

We remind the reader that the UK consists of four nations with an overall population of 64.5 million: England (54.3 million), Scotland (5.3 million), Wales (3.1 million) and Northern Ireland (1.8 million). In 1997, Scotland and Wales held referendums which determined in both cases the desire for increased self-government (along with Northern Ireland and the 1998 Good Friday Agreement), creating national assemblies to which a variety of powers – in particular, education – were devolved from the UK Parliament. Thus, we now have an educational policy ecosystem in the UK that was historically ruled by one parliament but now comprising three devolved assemblies responsible for four separate education systems.

In the context of increasing international focus on curriculum and qualification reform to support computer science education and digital skills in schools, the four education systems of the UK have proposed and implemented a variety of changes [9, 34, 5, 8], particular in England [6], with a new compulsory computing curriculum for ages 5-16 from September 2014 [12]. For universities across the UK offering computer science degrees [32], this school curriculum reform has had uncertain (and emerging) impact on the delivery of their undergraduate programmes, with the diversity of the educational background of applicants likely to increase before it narrows: it is certainly possible now for prospective students to have anywhere from zero to four or more years experience (and potentially two school qualifications) in computer science with some knowledge of programming.

Since 2012, there has been increasing scrutiny of the quality of teaching in UK universities, partly linked to the current levels – and potential future increases – of tuition fees (generally paid by the student through government-supported loans), as well as relative levels of graduate employability and the perceived value of professional body accreditation by industry. In February 2015, the UK (but in this respect only responsible for England) Department of Business, Innovation & Skills initiated independent reviews of science, technology, engineering and mathematics (STEM) degree accreditation and graduate employability,[1] with a specific focus – the Shadbolt review [36] – on computer science degree accreditation and graduate employability, reporting back in May 2016. Alongside a number of recommendations to address the apparently relatively high unemployment rates of computer sciences graduates, particular on the quality of data, course types, gender and demographics, the Shadbolt review split generalist universities in England into three bands ("low", "medium", "high"), based on the average (across all subjects) entrance tariff of incoming undergraduates (as defined by the national UCAS Tariff[2]); we have followed the same tariff banding during our analysis of the English results, so our data should allow comparisons.

Thus, in this evolving environment of new policies and curricula, as well as the emerging demands of innovative pedagogies and high-quality learning and teaching for computer science degree programmes, we present the findings from the first national scale survey of introductory programming languages at UK universities, providing a baseline for deeper analysis of the art, science and engineering of programming. Through this first UK-wide survey of universities, we identify and analyse trends in student numbers, programming paradigm, programming languages and environment/tools used, as well as the underpinning rationale for their selection.

## 2 Methodology

### 2.1 Recruitment of Participants

To recruit for the survey, a general call for participants was sent out to the Council of Professors and Heads of Computing (CPHC) membership; CPHC[3] is the representative body for university computer science departments in the UK, with nearly 800 members at over 100 institutions. The survey was hosted online from mid-May until the end of

---

[1] https://www.gov.uk/government/collections/graduate-employment-and-accreditation-in-stem-independent-reviews
[2] https://www.ucas.com/advisers/guides-and-resources/tariff-2017
[3] https://cphc.ac.uk

18-3

**An Analysis of Introductory Programming Courses at UK Universities**

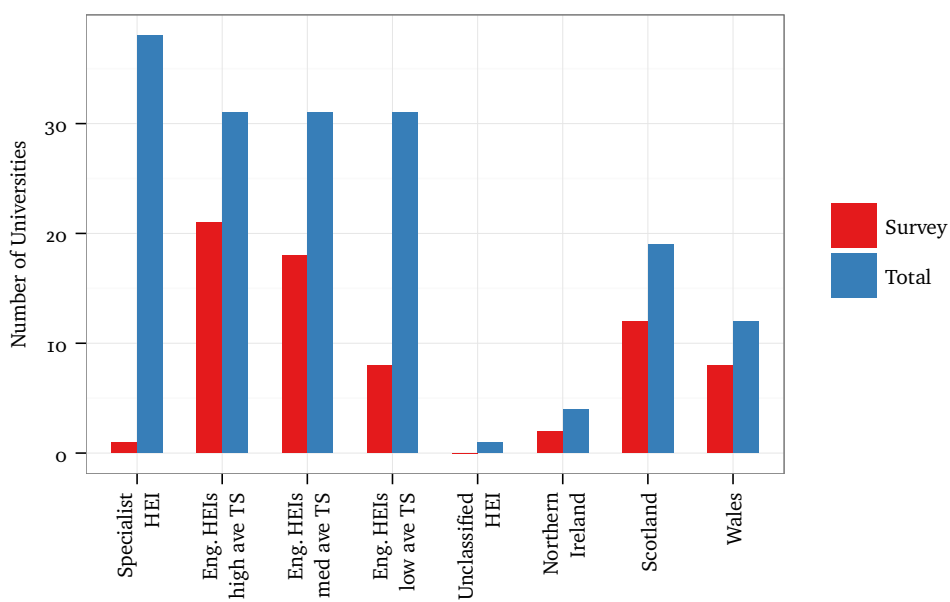

**Figure 1** The number of responding universities per Nation/Tariff Group

June 2016, with the invitation asking for the survey to be passed on to and completed by the most appropriate person in that institution. Due to the recruitment method, there were a number of duplicate responses from certain departments, and these were reconciled by direct enquiry.

The questions used in the survey were generously provided by the authors of the 2013 Australia and New Zealand survey [23], so as to allow direct comparison with the results of this survey. Where possible, questions were left unchanged, although a small minority were edited to reflect the UK target audience. As defined in the 2013 Aus/NZ survey, the terminology "course" was used for "*the basic unit of study that is completed by students towards a degree, usually studied over a period of a semester or session, in conjunction with other units of study*".

The first section of the survey asked about the programming language(s) in use, the reasons for their choice, and their perceived difficulty and usefulness. Then, questions regarding the use of environments or development tools; which ones were used, the reasons for their choice and the perceived difficulty. General questions about paradigm, instructor experience and external delivery were asked, along with questions regarding students receiving unauthorised assistance, and the resources provided to students. Finally, participants were asked to identify their top three aims when teaching introductory programming, and were also allowed to provide further comments.

In the 2013 Aus/NZ survey, participants were asked to rank the importance of the reasons for choosing a programming language, environment or tool. Due to technical limitations in the online survey tool used, it was not possible to do so in this survey, so Figure 3 only reports counts. Most questions were not mandatory; the exceptions were

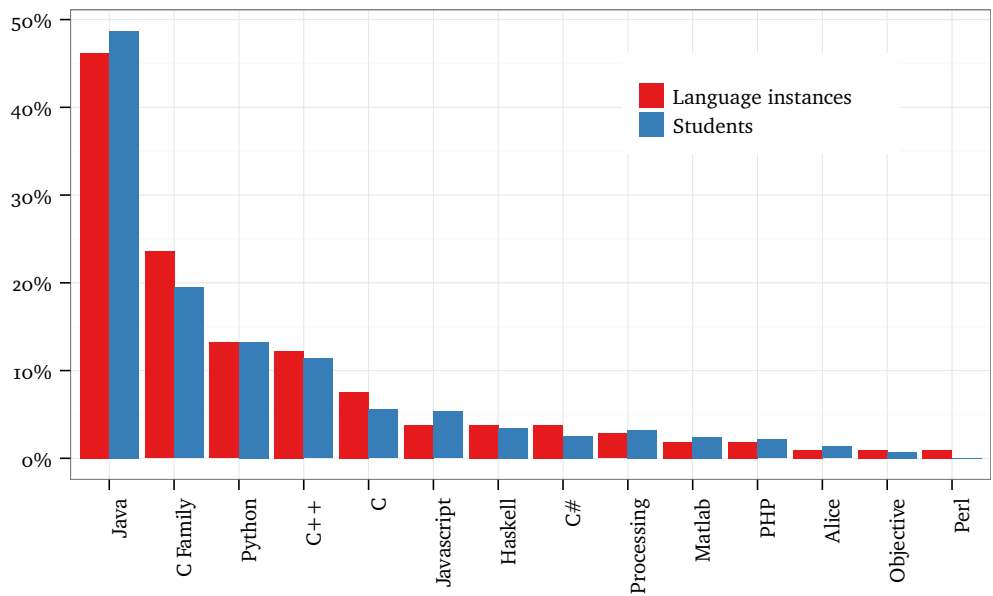

**Figure 2** Language popularity by percentage of courses and students (excl. The Open University)

"*what programming language(s) are in use?*" and a small number of feeder questions to allow the survey to function correctly.

## 3 Results

### 3.1 Universities and Courses

Upon completion of the survey, 155 instructors had, at least, started the survey; 61 of these dropped out before answering the mandatory questions, and a further 14 were duplicates. Therefore, the results presented here are drawn from the responses of 80 instructors from at least 70 institutions. Some participants did not answer all questions and thus the response rate varies by question.

Excluding the Open University's 3200 students, the participants in the survey represented 13462 students, with a mean of 173 (but a standard deviation of 88). Looking at Figure 1 we see good response rates, apart from the specialist higher education institutions (most of whom do not teach computing) and the "low tariff" English ones. Fewer of these teach computing; this factor alone explains the response rate. In Northern Ireland, we had responses from the two universities, but not the university colleges, which are historically initial teacher education colleges.

**An Analysis of Introductory Programming Courses at UK Universities**

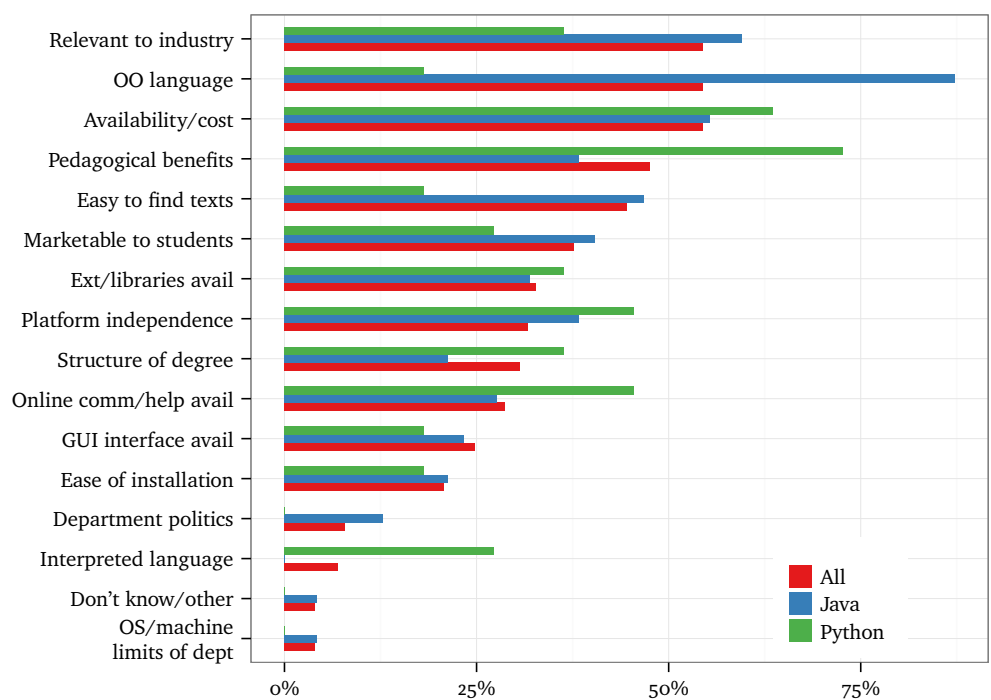

**Figure 3** Reasons given for choosing a programming language by percentage for: all languages; Java; and Python

### 3.2 Languages

A primary focus of this survey was to identify the programming languages in use in introductory programming courses. Participants were asked to select languages from a list of 22 programming languages and also had the option to choose "*Other*" and specify a language not included in the list. The majority of courses surveyed (59 out of 80, 73.8%) use only one programming language, with 17 using two (and only three and one institutions using three and four languages respectively). From the 80 courses, the total number of *language instances* is 106, as some courses use more than one language to teach introductory programming.

Of the 22 languages provided, 13 were selected at least once. The relative popularity of languages is shown in Figure 2, where the prevalence is given by the percentage of a language over all language instances (106 total), and weighted by student numbers (16662 total) per language instance. The programming languages that were not selected at all were: Actionscript, Ada, Delphi, Eiffel, Fortran, jBase, Lisp, Ruby and Visual Basic.

The relative popularity of languages is the immediate major difference with the 2013 Aus/NZ survey; their survey showed a dead heat (27.3% of language instances) between Java and Python, with Python winning (33.7% to 26.9%) when weighted by the number of students enrolled on the course. Our findings in Figure 2 show that Java is a clear winner by any metric, being used in over half the courses (61.3%) and just under half of all language instances (46.2%), while the runner-up, Python, is in use in 17.5% of courses and makes up 13.2% of language instances. The C family (C, C++ and C#)[4] together is in use in 31.3% of introductory programming courses, and scores 23.6% of language instances and 19.5% by students. Figure 2 also shows the effect of student-number weighting *but* we have excluded the Open University from this weighting, as its 3200 students learning Python (and Sense, a variant of Scratch) would have distorted the comparison.

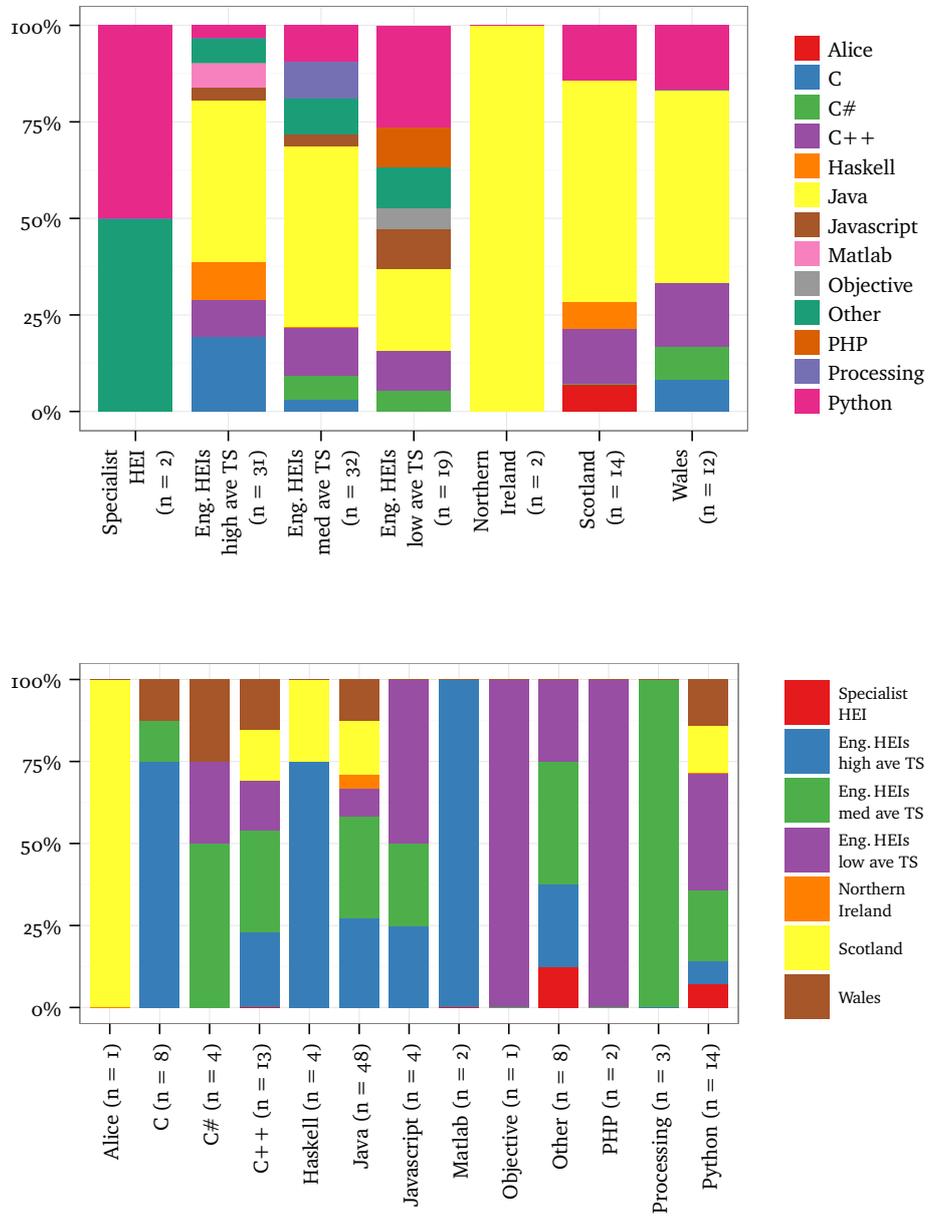

**Figure 4** The breakdown of programming languages by Nation and Tariff Groups

For each language selected, participants were asked to give the reasons for choosing that language for the introductory programming course. Figure 3 shows the frequency of these reasons for all languages grouped together and for Java and Python individually. When the reasons given are combined for all languages, three reasons tie for first place: *"relevance to industry"*; *"object-oriented language"*; and *"availability and cost to students"*, all chosen by 54.5% of participants who answered this question.

Looked at individually, the most popular reason given for choosing Java is *"object-oriented language"* at 87.2%, while Python scores highest on *"pedagogical benefits"*, at 72.7%. This may explain the popularity of Java: Java scores higher on *"relevance to industry"* and, perhaps somewhat surprisingly, much higher on *"object-oriented language"* than Python, which only scores 18.2%.

Figure 4 breaks down the choice of language by nation and tariff group. It is noticeable that the three English tariff groups differ significantly, with Python outnumbering Java in the low tariff universities, and C being almost exclusively in the high tariff universities. Figure 5 gives the instructors' views on languages. It is noteworthy that Java is among the most difficult, and not among the pedagogically most useful.

For each language chosen, instructors were asked whether the language was used: for the whole of the first programming course; for the first part of the first programming course, followed by another; after another language in the first programming course. Of 93 language instances, the majority (65%) are used for the whole of the introductory programming course, 14.0% of language instances are used in the first part of a course and 21.5% of language instances are used after another programming language; results are displayed in Figure 6.

### 3.3 Paradigm Taught

Instructors were asked which paradigm was being taught in their introductory programming course, regardless of what is traditionally thought to apply to the language(s) in use. This question, understandably, caused some dissatisfaction in the comments section, with many participants noting that more than one paradigm is taught in their course. Although this was to be expected, we wanted to be able to directly compare our results to the 2013 Aus/NZ survey, and so did not alter the question. The most popular paradigm is object-oriented with 50% (N = 40, 50%)

---

[4] One referee queried whether C# counts as "C family", describing it as *"much closer to Java"*. One can find apparently authoritative statements in both camps from the language designers of Java and C#. Further analysis of the four C# instances shows four different patterns: C# only; a wide range of languages; C++ followed by C# and Java followed by C#.

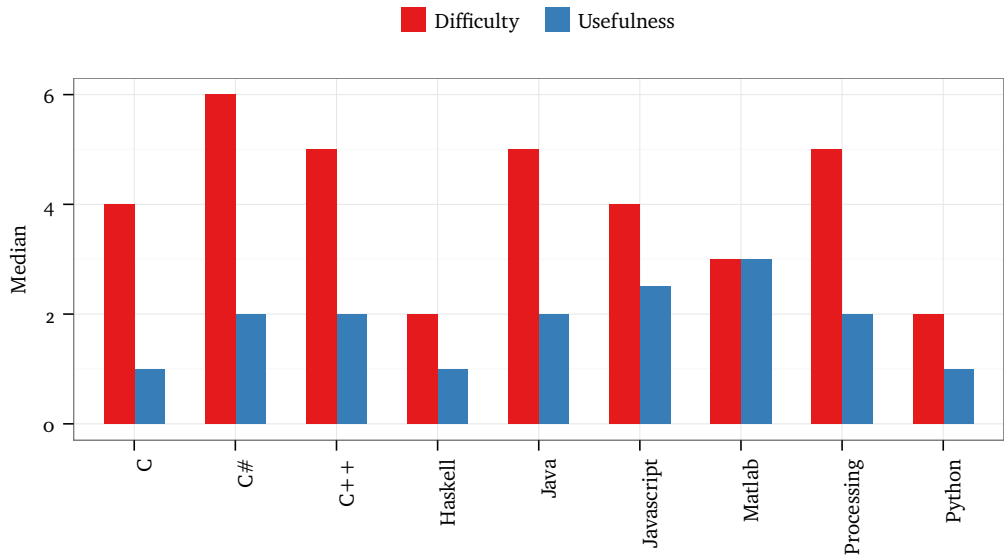

**Figure 5** The median of the perceived difficulty and [pedagogic] usefulness of language, where 1 is '*extremely easy/useful*' and 7 is '*extremely difficult/useless*'

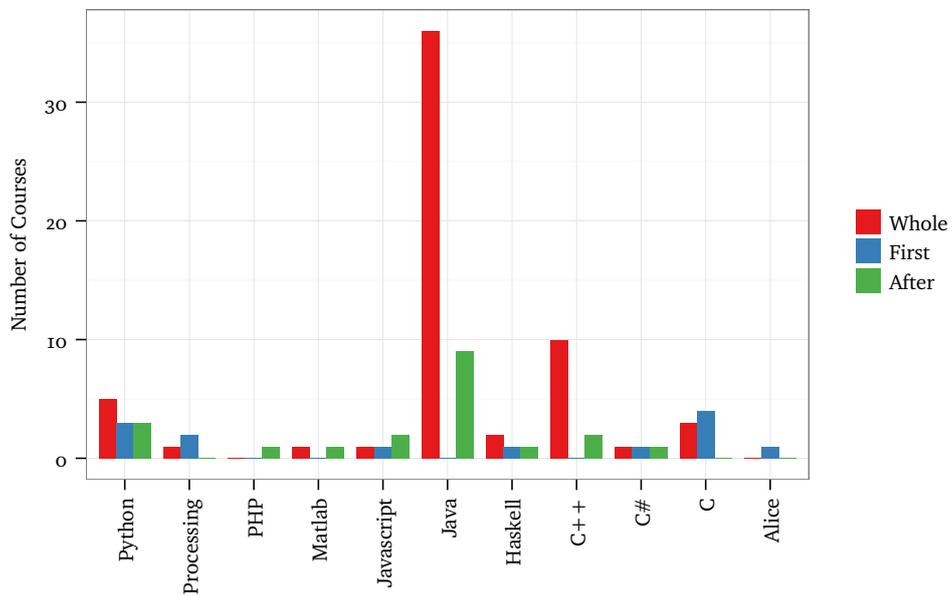

**Figure 6** For each language, whether the language is used: for the whole of the first programming course; for the first part of the first programming course, followed by another; after another language in the first programming course

**An Analysis of Introductory Programming Courses at UK Universities**

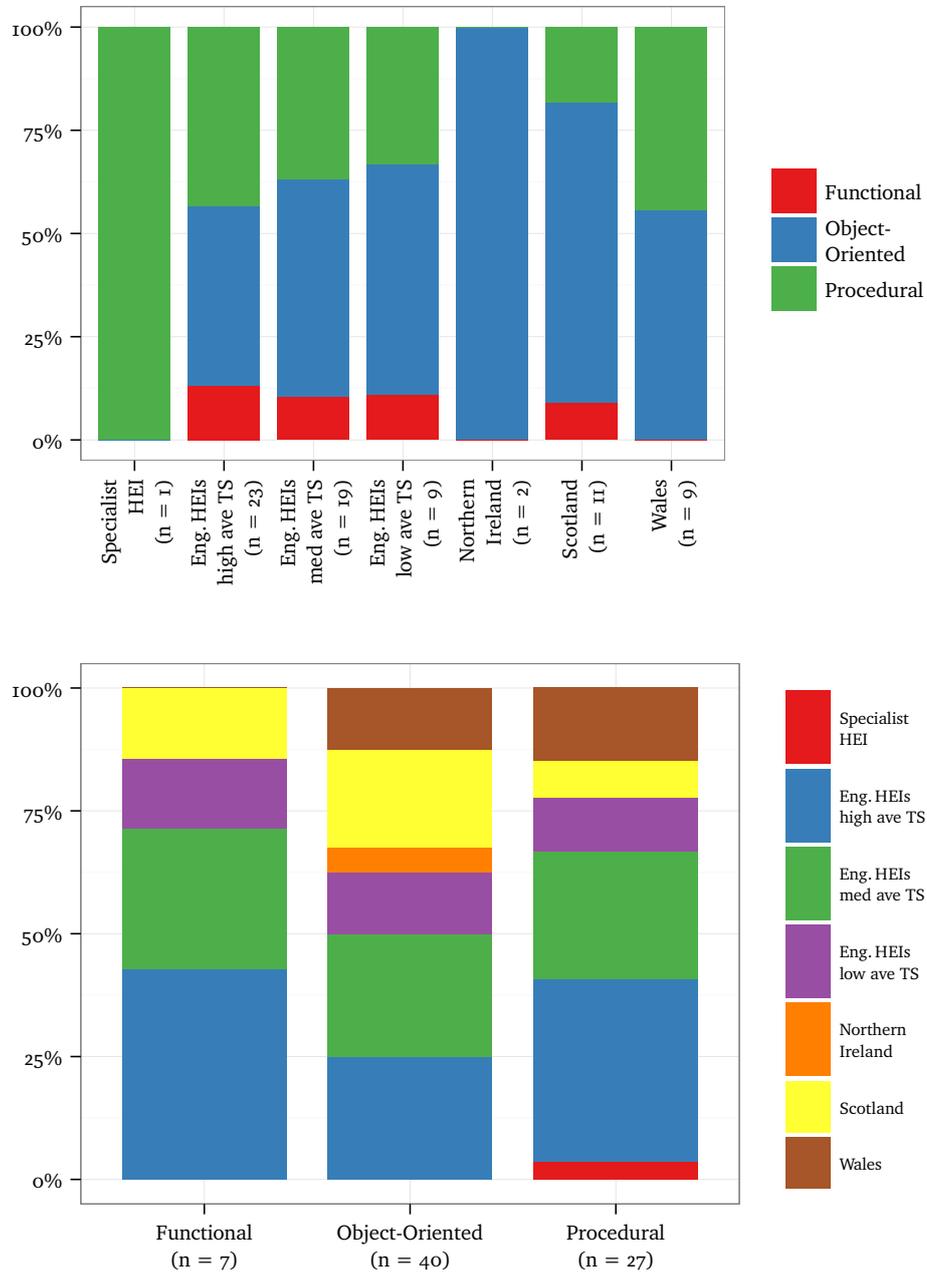

**Figure 7** The breakdown of the main paradigm in use for every Tariff Group

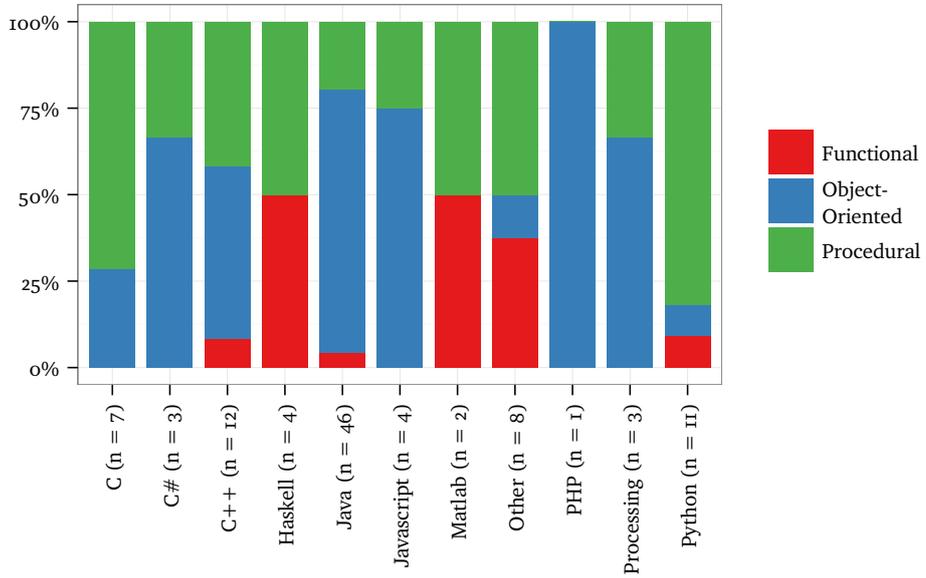

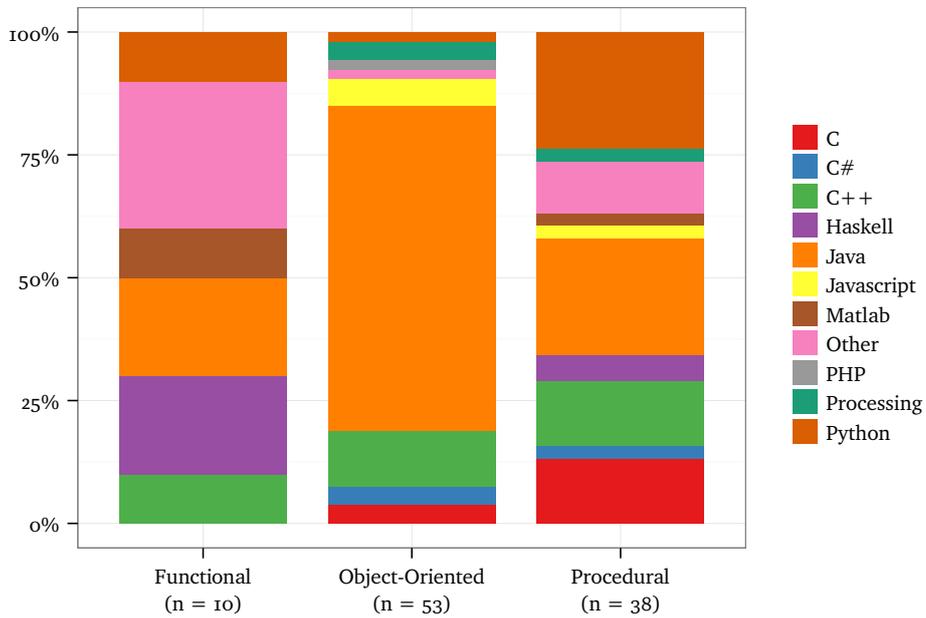

**Figure 8** The breakdown of the main paradigm in use for each programming language

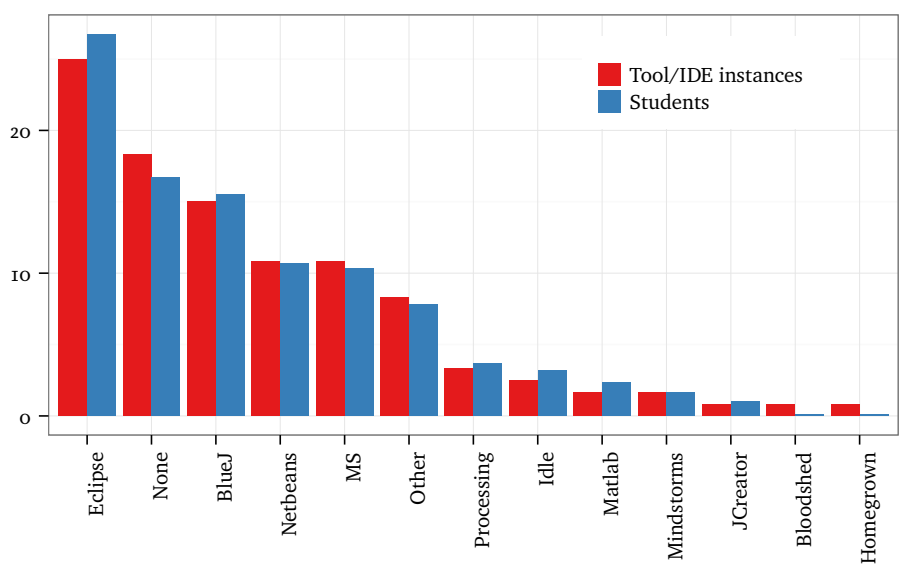

**Figure 9** Tool or environment popularity by percentage of courses and students

followed by procedural (N = 27, 33.8%) and functional (N = 7, 8.75%); logical was also offered as a choice but was not selected.

The results of the previous question were used to analyse – see Figure 7 – the prevalence of paradigms across nations and tariff score groups. Caution must be applied when interpreting these results, as participants could only choose one paradigm, even though more may be in use.

In the same way as above, the languages chosen were analysed – see Figure 8 – with regard to the main paradigm in use. Again, caution must be applied, as for a given course, only one paradigm is chosen, even though more than one language and/or paradigm may be in use. This explains the respondents who used C, but stated that object-oriented was the main paradigm, for example. More surprising is the fact that Python was almost exclusively viewed as procedural.

## 3.4 Instructor Experience

Participants were asked: *"How many years have you been involved in teaching of introductory programming?"*. The results, shown in Table 1, indicate that of the survey participants, the average was between 10-20 years.

**Table 1** The number of years the instructor has been teaching introductory programming.

| Years | <2 | 2 - 5 | 5 - 10 | 10 - 20 | 20 - 30 | >30 |
|---|---|---|---|---|---|---|
| Instructors | 3 | 9 | 9 | 27 | 19 | 7 |

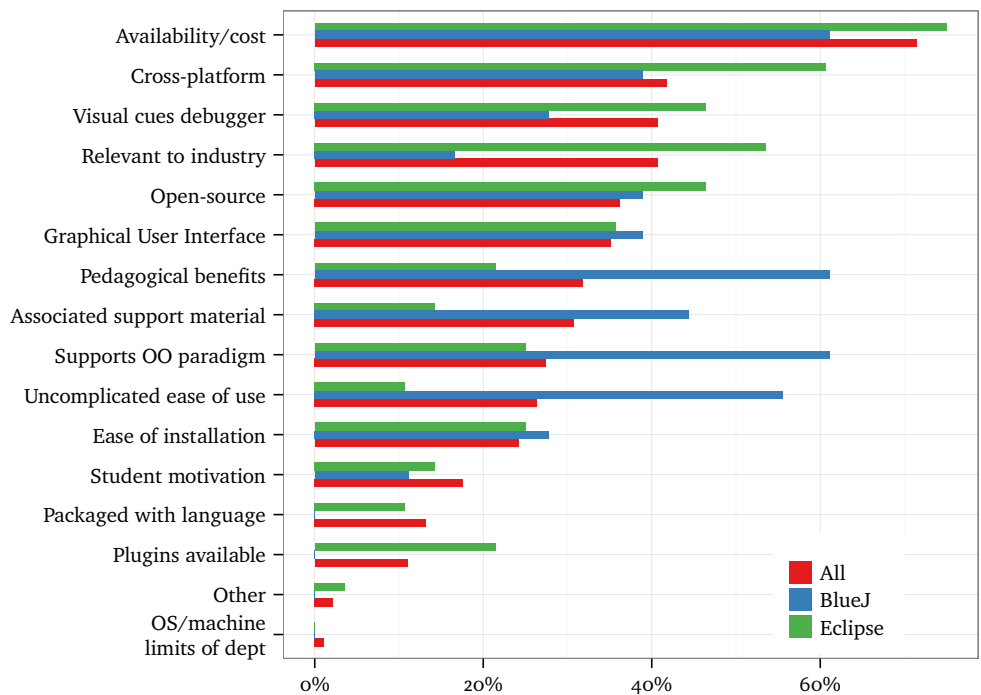

**Figure 10** Reasons given for choosing a tool or environment by percentage for: all tools and environments; BlueJ; and Eclipse

### 3.5 IDEs and Tools

Participants in the survey were asked if they encouraged students in the first programming course to use environments and/or tools beyond simple text editors and command line compilers. The majority of participants of this question (74.4% of 78 instructors) responded that they did encourage tools. Of the 58 instructors that did select a tool/IDE, the majority (58.6%) use only one, with 25.8% and 10.3% using two and three respectively; very few (5.2%) used four or more, with one respondent using eight.

The survey asked participants to select the tools and IDEs in use in their introductory programming course out of a list of 24, with the option to specify "*Other*". Of the 24 provided, 12 were chosen at least once. The relative popularity of IDEs and tools is shown in Figure 9. The most popular tool/IDE in the survey was Eclipse, reported in 37.5% of courses and scoring 25.0% of tool/IDE instances, and 26.8% when weighted by students. Following this is "*No Tool/IDE*", which accounts for 27.5% of courses. The second most popular tool/IDE is BlueJ, which was reported in 22.5% of courses and scored 15.0% of tool/IDE instances, and 15.5% when weighted by students. Participants were also asked why each tool/IDE was chosen for their course, and asked to select from a list of reasons. The results of this are give in Figure 10, for all tools and IDEs grouped together, and for the two most popular choices, Eclipse and BlueJ. The tools and IDEs not selected at all were: AdaCore, Alice, App, Browser, Greenfoot, Jeroo, Jython, KTechLab, MySQL, Pelles, Quincy, Wing101 and Xcode.

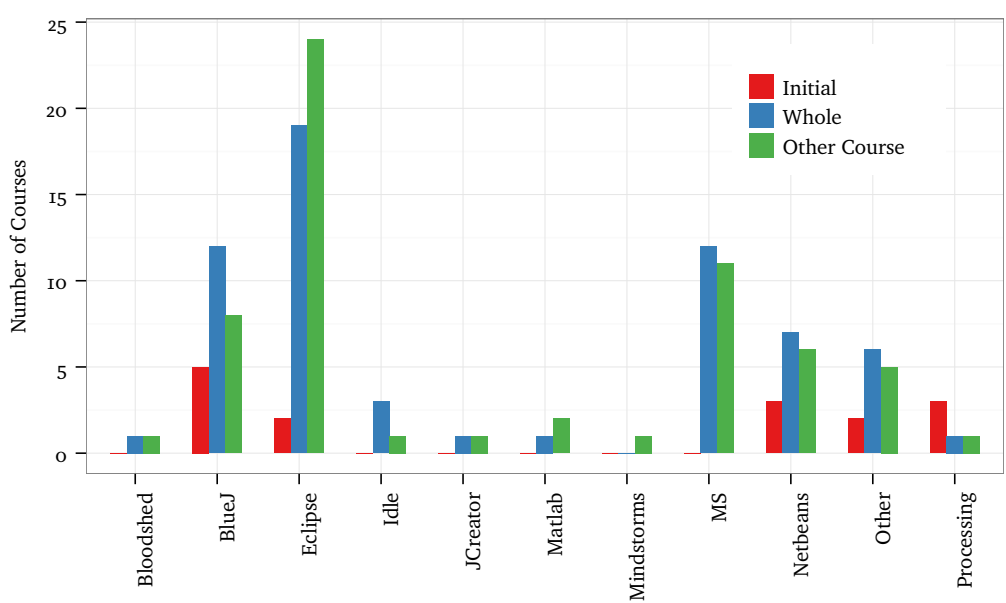

**Figure 11** For each tool or environment, whether it is used: for an initial part of the first programming course; throughout the whole of the first programming course; in any other course in the degree

### 3.5.1 Reuse of Tool/IDE

Instructors were also asked whether the tool/IDE was used for an initial part of the first programming course or throughout the whole of the course; and whether it was used in any other course in the degree (Figure 11).

### 3.5.2 Difficulty of Tool/IDE

In addition to this, instructors were asked to rate how difficult *they* found the tool/IDE on a Likert scale from "*Extremely Easy*" (1) to "*Extremely Difficult*" (7), and also how difficult they believed *the students* found the tool/IDE, shown in Figure 12.

We note that, while Eclipse is the most popular tool by some way, it is also deemed to be most difficult. This, apparently perverse, practice might be explained by the extent of re-use of Eclipse in other courses.

## 3.6 Other Aspects of the Course

### 3.6.1 External Delivery

Participants were asked "*Do you offer external delivery of your course? (i.e. do you have options for your course where students are not required to attend regular lectures, workshops, labs or tutorials?)*". The responses to this question were overwhelmingly in the negative; 70/74 (94.6%) answered "*No*".

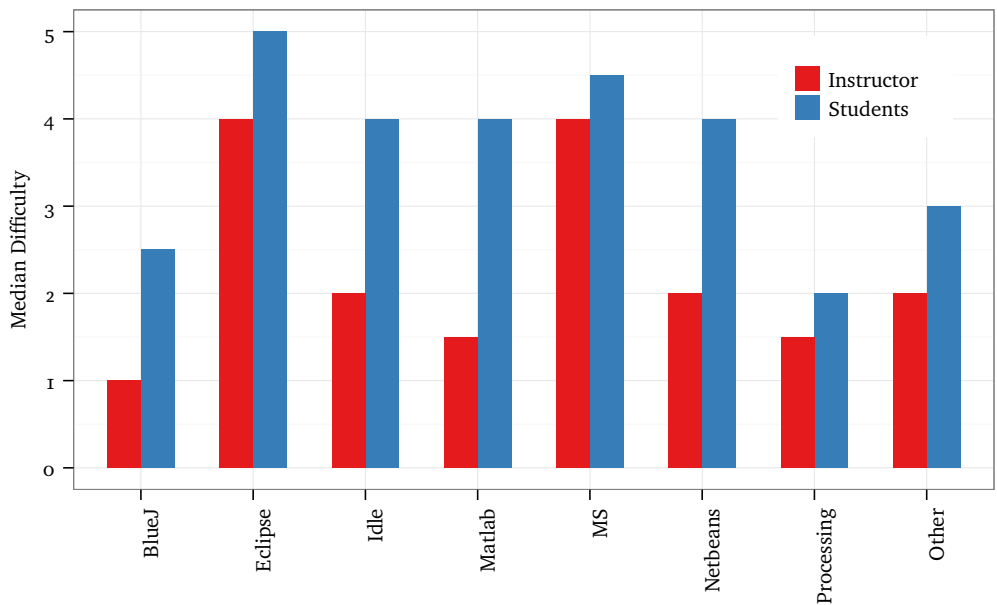

**Figure 12** The median difficulty rating of tool for the instructor and students to use, where 1 is *extremely easy* and 7 is *extremely difficult*

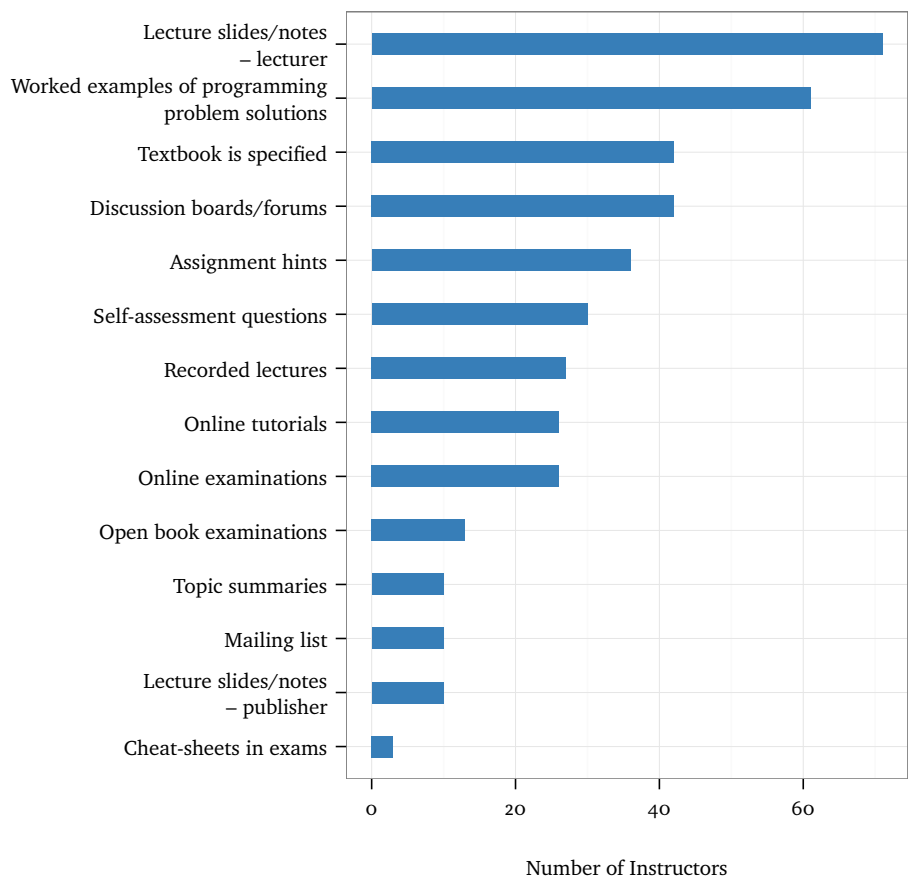

**Figure 13** Resources provided to students

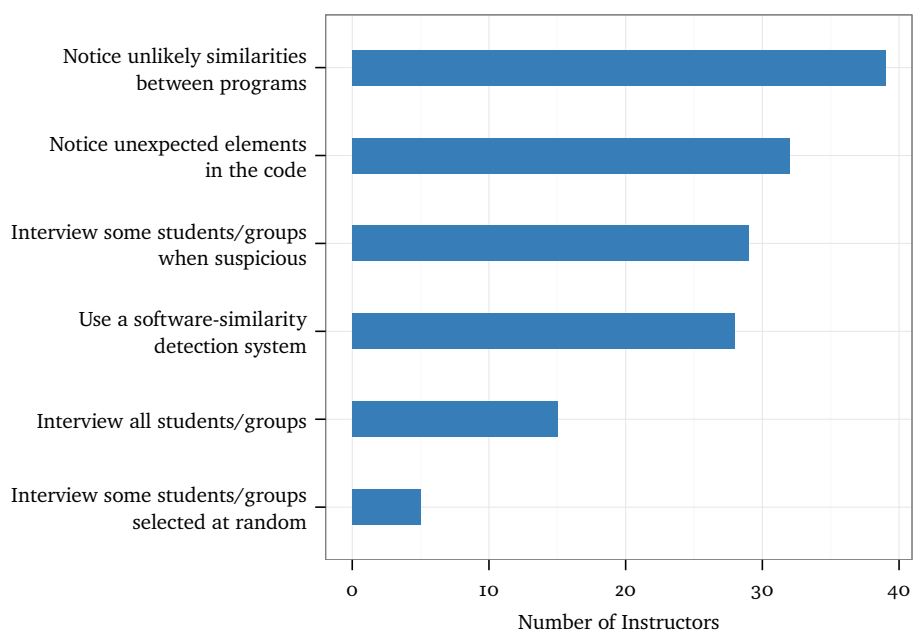

**Figure 14** Steps taken to determine whether students have received unauthorised assistance on assignments

**3.6.2 Resources Provided to Students**

The questionnaire asked about the resources in terms of examples, books etc. provided to students. The results are rather similar to the 2013 Aus/NZ survey [23, Figure 14] and are displayed in Figure 13. The most popular resources selected were: "*lecture slides or notes provided by the lecturer*" in first place, "*worked examples of programming problems/solutions*" in second, and third place was shared by "*textbook is specified*" and "*discussion boards/forums*".

**3.6.3 Unauthorised Assistance**

The vast majority of instructors surveyed (89.3%) do consider the possibility that students or groups of students may be receiving unauthorised assistance (e.g. from other students in the class, from people outside the class, or via the internet) when doing assignments. When asked how concerned they were about this possibility, 9 answered "*not concerned*", 39 answered "*somewhat concerned*" and 17 reported "*very concerned*".

We also asked participants: "*What steps do you take to try to determine whether students have received unauthorised assistance on assignments?*". The details are displayed in Figure 14 and range from "*notice unlikely similarities*" (59.1% of the 66 instructors who responded to this question) to "*interview some students/groups at random*", selected by only five instructors.

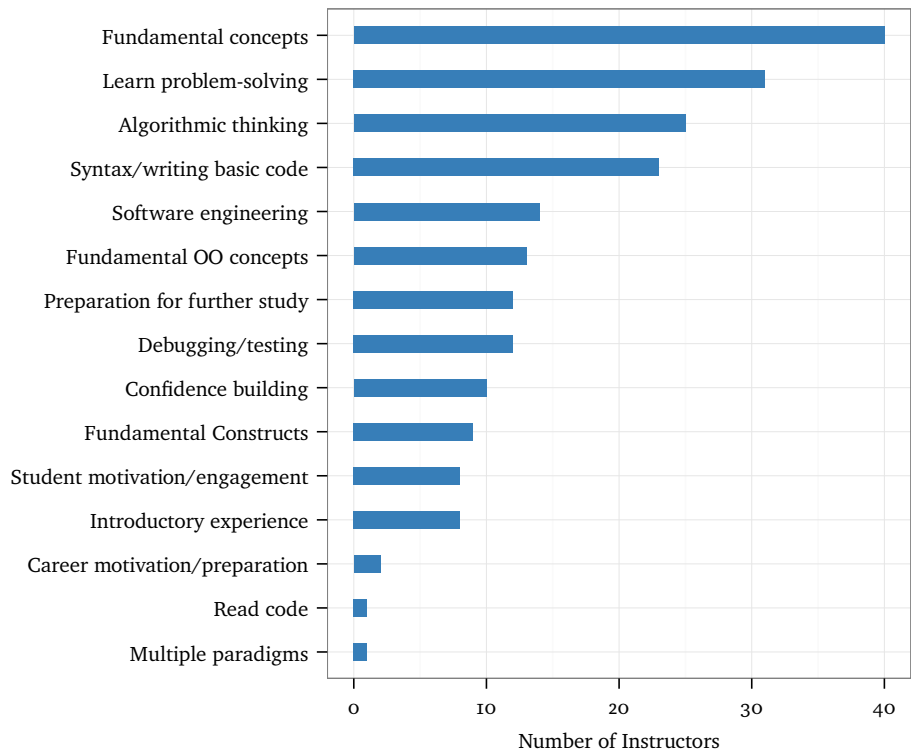

**Figure 15** Aims of the introductory course

### 3.7 Aims of an Introductory Programming Course

The 2013 Aus/NZ survey asked their respondents for the aims of their introductory programming course. They, and we, asked for (up to) three aims. The authors then attempted to classify the free-text answers into the same categorisation as [23] used. While it is trivial to map the written aim "*Thinking algorithmically*" to "*Algorithmic thinking*" [23] and so on, many were not so clear: for example, we mapped "*To learn a specific language*" to "*Syntax/writing basic code*". There were also a class of aims, such as "*Establish professional software development practices*", that seemed coherent, but did not map clearly to the [23] aims; these we have categorised as "*Software Engineering*". Results of this question are shown in Figure 15.

## 4 Discussion

### 4.1 Comparison with Australasian Survey

Here we compare with the latest Australasian survey [23]; we have already commented on the major difference in language choice, which colours many of the other comparisons. In fact, the UK's language choices seem more similar to Australasia's 2010 choices [24] and [23, Table 4] than even Australasia's 2013 choices. It is hard to know which comes first, but we also notice that our difficulty/utility data (Figure 5) is somewhat different from [23, Figures 7/8].

Another difference in the tools/environments used is demonstrated by Figure 9 versus [23]'s Figure 11. There, "*None*" and "*Other*" were the top two categories, with IDLE, at 15%, the most popular named product. In the UK, "*None*" is second, "*Other*" is sixth and IDLE eighth. Eclipse, the UK favourite, was an "also ran" in [23].

### 4.2 The UK Context

As presented in Section 3.2, our findings show that Java is the most popular introductory programming language in UK universities, more than twice as popular as Python in second place; the C family of languages (C, C++ and C#) together is in use in nearly a third of introductory programming courses. We were surprised by the viewpoint expressed of Python, as a multi-paradigm language, as being largely procedural[5]; from the authors' experiences, the dominance of Java has been a trend for the past ten years, but we would expect to see a steady increase in Python due to influences from the changes to school curricula in the UK [6].

We note that from a smaller survey conducted in July 2014, Python is the most popular language for teaching introductory computer science courses at top-ranked US university departments; specifically, eight of the top 10 CS departments (80%), and 27 of the top 39 (69%), teach Python in introductory CS0 or CS1 courses [18]. This together with [23] and Figure 5 might make one question the UK's domination by Java, although longstanding industry popularity as measured by community indices may still be a significant determining factor [40, 38].

From a UK education policy perspective, a new national Teaching Excellence Framework has been proposed, with a core ambition to "*to raise the quality and status of teaching in higher education institutions*"; excellence is to be measured through a series of proxy metrics that include student satisfaction, retention and graduate employability.[6] There have been significant sector concerns about the aims of the framework – as well as the statistical rigour of the metrics – more so in the context of it being used for benchmarking[7] "teaching excellence" (particularly as the TEF will not yet be conducted at the individual subject level, but at the institutional level), as well as deciding whether institutions are allowed to raise tuition fees in the future. It remains to be seen how this will affect undergraduate computer science degree curricula in UK institutions going forward, especially if there is renewed demand for meeting the immediate (but potentially transient) demands of the IT industry with specific tools, languages and environments, as well as reformed professional body accreditation as per the 2016 Shadbolt review [36].

---

[5] We can only speculate why this is; one reason could be the nature of many of the texts available: see [27], and for example a popular freely-available Python text [14] which the third author has used while teaching teachers introduces classes only in chapters 15-17 (of 19 in total).
[6] http://www.hefce.ac.uk/lt/tef/
[7] Many UK newspapers produce "University League Tables", all based on much the same published data. The new Teaching Excellence Framework will grade universities as bronze/silver/gold, and it seems inevitable that the newspaper league tables will use these in their league tables.

The UK's Higher Education Academy – the national body which champions teaching quality – has previously supported initiatives for improving learning and teaching in computer science, including innovative pedagogies for programming [7, 11], but we have not yet seen the necessary development of sustainable discipline-specific communities of practice, both at the local and national level, to capture and share best practice.[8]

## 5  Future Work

This national survey provides valuable context for better understanding of the role and effectiveness of programming education in UK universities. Furthermore, how this impacts more broadly across the education pipeline: through significant curriculum reform, as well as scrutiny of the effectiveness of pedagogies for teaching principles of programming and software engineering (in essence: software carpentry, providing the knowledge, skills and understanding to create useful and usable software for a variety of domains). Moving forward will require an mixed economy of rigorous pedagogical research, as well as the application of personal experiences of languages, tools, environments, models and styles. Only through this blend of the art, science and engineering of programming will we see significant steps towards improving programming (and thus computer science) education in the UK.

### Acknowledgements

The authors would like to thank the participants for their engagement with the survey, as well as R. Mason and G. Cooper from Southern Cross University, Australia, for providing us with their survey and permission to use it. We are grateful to the GW4 Alliance (Universities of Bath, Bristol, Cardiff and Exeter) for funding the survey and Alan Hayes (University of Bath) for group coordination.

Those data created during this research project that do not infringe the anonymity of the respondents are openly available from the University of Bath data archive at http://doi.org/10.15125/BATH-00246.

---

[8] However, a new initiative announced by the Royal Society in 2016 is aiming to addresses this in UK schools: https://royalsociety.org/topics-policy/projects/computing-education/

**About the authors**

**Ellen Murphy** is a Commercial Research Associate at the Bath Institute for Mathematical Innovation, University of Bath. Contact: e.murphy@bath.ac.uk

**Tom Crick** is Professor of Computer Science & Public Policy at Cardiff Metropolitan University. Contact: tcrick@cardiffmet.ac.uk and @ProfTomCrick

**James H. Davenport** is Hebron & Medlock Professor of Information Technology at the University of Bath. Contact: j.h.davenport@bath.ac.uk and @JamesHDavenport